\documentclass[runningheads]{llncs}

\usepackage{floatrow}
\newfloatcommand{capbtabbox}{table}[][\FBwidth]

\usepackage{titlesec}
\usepackage{todonotes}
\usepackage{fixltx2e}
\usepackage{amssymb}
\usepackage{graphicx}
\usepackage{wrapfig}
\usepackage{subfig}
\usepackage{subfloat}
\usepackage{url}
\usepackage{amsmath}
\usepackage{verbatim}
\usepackage{multirow}
\usepackage{paralist}
\usepackage{mathptmx}
\usepackage[inline]{enumitem}
\usepackage{times}
\usepackage{dirtree}
\usepackage{tabularx}
\usepackage{tikz}
\usetikzlibrary{arrows,%
                calc,%
                chains,%
                decorations.pathmorphing,%
                decorations.pathreplacing,%
                decorations.shapes,%
                graphs,%
                matrix,%
                positioning,%
                shapes.geometric,%
                shapes,%
                shapes.symbols,%
                trees
   			  }
\usepackage{array}
\usepackage{booktabs}
\usepackage{graphicx}
\usepackage{pgfplots}
\usepackage{floatrow}

\usetikzlibrary{automata,arrows}
\usetikzlibrary{datavisualization}
\usetikzlibrary{datavisualization.formats.functions}
\usetikzlibrary{positioning, calc, backgrounds, arrows}
\usetikzlibrary{shapes.geometric}
\usetikzlibrary{decorations.pathreplacing}
\usetikzlibrary{patterns}

\let\olddefinition\definition
\def\definition{\olddefinition\footnotesize}

\newcommand{\optional}[1]{}

\usepackage[ruled,vlined,boxed,linesnumbered]{algorithm2e}
\SetKwRepeat{Do}{do}{while}
\SetKwInOut{Input}{input}
\SetKwInOut{Output}{output}

\graphicspath{{images/}}

\titlespacing\section{0pt}{9pt plus 2.0pt minus 2.0pt}{9pt plus 2.0pt minus 2.0pt}
\titlespacing\subsection{0pt}{9pt plus 2.0pt minus 2.0pt}{9pt plus 2.0pt minus 2.0pt}
\titlespacing\subsubsection{0pt}{9pt plus 2.0pt minus 2.0pt}{9pt plus 2.0pt minus 2.0pt}

\begin{document}

\title{Five guidelines to improve context-aware process selection: an Australian banking perspective}
\author{Nigel Adams\inst{1} \and
            Adriano Augusto\inst{1} \and
		    Michael Davern\inst{1} \and
			Marcello La Rosa\inst{1}
			 }
\titlerunning{ }
\authorrunning{ }

\institute{
           University of Melbourne, Australia\\
           naadam@student.unimelb.edu.au\\
   		    \{a.augusto, m.davern, m.larosa\}@unimelb.edu.au 
}

\maketitle 

\begin{abstract}
As the first phase in the Business Process Management (BPM) lifecycle, process identification addresses the problem of identifying which processes to prioritize for improvement. Process selection plays a critical role in this phase, but it is a step with known pitfalls. Decision makers rely frequently on subjective criteria, and their knowledge of the alternative processes put forward for selection is often inconsistent. This leads to poor quality decision-making and wastes resources. In recent years, a rejection of a one-size-fits-all approach to BPM in favor of a more context-aware approach has gained significant academic attention. In this study, the role of context in the process selection step is considered. The context is qualitative, subjective, sensitive to decision-making bias and politically charged. We applied a design-science approach and engaged industry decision makers through a combination of research methods to assess how different configurations of process inputs influence and ultimately improve the quality of the process selection step. The study highlights the impact of framing effects on context and provides five guidelines to improve effectiveness.

\end{abstract}

\section{Introduction}\label{sec:introduction}
Business Process Management (BPM) is concerned with managing how organizational processes are executed to deliver consistent and improved services and products to customers~\cite{dumas2018fundamentals}. It is both well-researched and an area of increasing importance in practice~\cite{vom2016role}. However, implementation success can be elusive~\cite{trkman2010critical}. In response, there has been significant academic effort focused on identifying the critical success factors (CSFs) and principles associated with a BPM implementation~\cite{trkman2010critical,vom2014ten,malinova2018identifying}. 

The first phase in the BPM lifecycle is process identification~\cite{dumas2018fundamentals}. Given an organization is unlikely to have the resources to improve all business processes, this phase addresses the problem of identifying which processes to prioritize~\cite{dumas2018fundamentals}. Process selection plays a critical role in this phase, but it is a step with known pitfalls~\cite{malinova2018identifying}. The authors in~\cite{dumas2018fundamentals} introduce a process portfolio selection tool, prioritizing processes based on three criteria: \emph{Strategic Importance}, \emph{Health} and \emph{Feasibility}. However, along with other proposed selection criteria~\cite{davenport1993process,elzinga1995business,leshob2018towards}, the information required to quantify these criteria is frequently difficult to source, qualitative and subjective~\cite{archer1999integrated,cho2011study,elzinga1995business}. These issues are compounded by the fact that prioritization is a group decision-making process~\cite{elzinga1995business} and ``Only a handful of stakeholders have a full overview of all the business processes in an organization"~\cite{dumas2018fundamentals}. Hence Decision Maker (DM) familiarity with alternative processes is likely to be inconsistent.

Much of the process selection research focuses on the choice of evaluation technique to prioritize processes. The authors in~\cite{wkatrobski2019generalised} identified 56 different multi-criteria decision analysis methods. Other studies focus on identifying processes for a specific improvement method~\cite{becker1999identifying,wanner2019process,leshob2018towards}, or on the process of process/project selection~\cite{elzinga1995business,archer1999integrated}. 

However, there is little discussion of the role that context plays in process selection. The role of context in BPM more broadly is one that has received considerable academic attention in recent times~\cite{melao2000conceptual,rosemann2006context,rosemann2008contextualisation,vom2016role}. It has been considered in a range of settings: understanding processes and process modeling~\cite{melao2000conceptual}, context-aware process design~\cite{rosemann2006context}, method selection~\cite{vom2021context}, as a categorization technique~\cite{vom2018frameworks}, and as a BPM principle~\cite{vom2014ten}. 

In this study, we examine the role of context at the key stages of the process selection step~\cite{archer1999integrated}. We test our proposition that DMs with Context-aware Criteria (CaC) will make better, more logical decisions, where CaC are the common criteria identified in~\cite{dumas2018fundamentals} supported by relevant contextual factors. The study follows a design-science approach, leveraging the work in~\cite{adams2022banks} as the motivating case. Over four iterations, a total of 75 senior managers (i.e., DMs) from the Australian banking industry were asked to indicate their level of support for including either a core or a support process in the scope of a process mining project. For each iteration, we adapted the configuration of the process inputs based on the observed behavior and process metrics from the prior stage. The DMs were engaged in a combination of online surveys, focus groups, and a Delphi study. Our research questions were: 

\begin{itemize}
    \item \textbf{RQ1:} How do Context-aware Criteria influence process selection decision making?
    \item \textbf{RQ2:} How should a context-aware, process selection step be configured to improve decision-making quality and logic?
\end{itemize}

Our study exposes the reality of process selection in practice as both politically charged and subject to bias, and highlights the impact of framing effects on context. We contribute five guidelines for configuring the stages of process selection~\cite{archer1999integrated}, and identify areas for future research. 

The remainder of this paper is organized as follows. In Section 2, we summarize the related work. In Section 3, we introduce our study design and the outcomes of the assessments. We discuss our findings, the guidelines and limitations of the study in Section 4. In Section 5, we present our conclusions and directions for future research.

\section{Background and Related Work}\label{sec:background}
In this section, we provide a brief background on process selection within the BPM Lifecycle, and the role of context in BPM. Also, we introduce the extensively researched topic of prospect theory and framing effects required to understand our results. 

\subsection{Business Process Management and Process Selection}
As one of two steps in the process identification phase of the BPM Lifecycle, process selection is critical to the success of BPM~\cite{elzinga1995business}. Its intent is to help organizations prioritize a portfolio of processes to manage and improve~\cite{dumas2018fundamentals}. Much of the research focuses on identifying criteria and the choice of evaluation technique. Davenport applies strategic relevance, process health, process scope, and the political and cultural environment as criteria~\cite{davenport1993process}. Kettinger et al., map processes to critical success factors and rate processes as ``Essential" or ``Desirable"~\cite{kettinger1997business}, whereas Cho and Lee~\cite{cho2011study} align criteria to the four perspectives of the balanced scorecard, and the authors in \cite{leshob2018towards} evaluate against eligibility, potential and relevance. Dumas et al., state that the most commonly used criteria are strategic importance, process health and feasibility~\cite{dumas2018fundamentals}. 

There are a wide-range of techniques proposed in the literature to  prioritize process alternatives~\cite{wkatrobski2019generalised}. Examples include the Analytical Hierarchy Process (AHP)~\cite{saaty2008decision}, fuzzy AHP~\cite{cho2011study}, weighted criteria utility theory verified by expert opinion~\cite{terres2010selection}, fuzzy quality function deployment and goal programming~\cite{hakim2016fuzzy}, the TOPSIS method~\cite{darmani2013business}, and process mining~\cite{wanner2019process}. In terms of the selection process, Becker et al.~\cite{becker1999identifying} propose six ``knock-out" criteria as a filter in their weighted-scoring model, while Archer et al.~\cite{archer1999integrated} propose a five-stage framework for selecting a project portfolio.
 
Process selection is a difficult, multi-attribute, group decision-making process~\cite{elzinga1995business}. Dumas et al.~\cite{dumas2018fundamentals} state that the additional information required to support the criteria is frequently subjective and the performance metrics are not always readily available. Malinova and Mendling identify five pitfalls at the process identification stage and a lack of criteria leads to a default selection of core processes over support processes and/or management processes~\cite{malinova2018identifying}. 

\subsection{Context in BPM}
Recognizing that a one-size-fits-all model is prone to failure, the role of context in BPM has grown in importance~\cite{melao2000conceptual,rosemann2006context,vom2016role}. Mel{\~a}o and Pidd consider business processes from four different perspectives including one that aims to capture the socio-political elements~\cite{melao2000conceptual}. Benner et al.~\cite{benner2003exploitation} highlight the difference between exploitative and exploratory BPM, indicating the need for an ambidextrous organizational model. Rosemann and Recker consider the extrinsic factors that act as the stimulus for change and note the challenge of identifying contextual variables relevant to process design~\cite{rosemann2006context}. Subsequently, Rosemann and colleagues~\cite{rosemann2008contextualisation} propose the onion model as a taxonomy for context classification, whereas vom Brocke et al. propose a context framework organized around four contextual dimensions: the BPM goal, the business process, the organization, and the environment~\cite{vom2016role}. Academics have applied context-awareness in a range of BPM scenarios from context-aware process design~\cite{rosemann2006context}, to method selection~\cite{vom2021context}, and as a principle of good BPM~\cite{vom2014ten}. However, a recent review found that 70\% of BPM methods are context-independent~\cite{denner2018context}. 
 
\subsection{Prospect Theory and Framing Effects}
To theoretically conceptualize context, we turn to the idea of framing from Prospect Theory, developed by Kahnemann and Tversky as an alternative decision-making model to expected utility theory~\cite{kah1979prospect}. Prospect Theory was formulated to explain individual choice made under uncertainty~\cite{edwards1996prospect}. It is best known for claims that people overweight losses with respect to comparable gains, are risk averse with respect to gains, and risk acceptant with respect to losses, the relationship with probabilities is non-linear, and given the asymmetrical treatment of gains and losses framing a problem around a reference point has a critical influence on their choices~\cite{levy1996loss}. It was subsequently revised to accommodate limitations associated with stochastic dominance and handling prospects with a large number of outcomes~\cite{tversky1992advances}. Academics have taken two approaches, choice shift and choice reversal, to evaluate and replicate the framing effects in the original study, but with mixed success~\cite{druckman2001evaluating}. Criticism includes the difficulty in determining the reference point and questioning the validity of framing effects in complex decisions~\cite{list2004neoclassical}. Nonetheless, as our empirical work reveals, we find the concept of framing useful in understanding the role of context.

\section{Study Design and Outcomes}\label{sec:design}
In this section, we discuss the design and outcomes of our study. To test our proposition, we needed to establish a base case, and then assess the impact of different input configurations on decision quality and decision logic. Starting with a focus on decision logic, we adopted a design-science approach and iterated through four stages until we found a configuration that resulted in improvement in both metrics. Adapted from the study in~\cite{archer1999integrated}, and with a strategic goal held constant throughout the exercise, our configurable inputs were: i) the CaC; ii) the processes to be prioritized; and iii) the DMs (our addition to the inputs in~\cite{archer1999integrated}). The configuration choices we made at each stage were a direct result of the observed behavior and output metrics from the prior stage (Table \ref{tab:config_metrics}). The reasoning behind our design choices is captured in the stage summaries below, where we summarize the steps for each stage, demonstrate the quantitative and qualitative outcomes, and provide observations. It is important to note that, given the different configurations, the methods varied by stage. 

\begin{table}[hbpt!]
\centering
{\scriptsize{
\centering
\begin{tabular}{llcccc}
\hline
\multicolumn{1}{|l|}{\textbf{Outputs}}         & \multicolumn{1}{l|}{\textbf{Item}}    & \multicolumn{1}{c|}{\textbf{Stage 1}} & \multicolumn{1}{c|}{\textbf{Stage 2}} & \multicolumn{1}{c|}{\textbf{Stage 3}} & \multicolumn{1}{c|}{\textbf{Stage 4}} \\ \hline
\multicolumn{1}{|l|}{Metrics}        & \multicolumn{1}{l|}{Decision Quality}     & \multicolumn{1}{c|}{N/A}              & \multicolumn{1}{c|}{No}               & \multicolumn{1}{c|}{Yes}              & \multicolumn{1}{c|}{Yes}              \\ \hline
\multicolumn{1}{|l|}{}                        & \multicolumn{1}{l|}{Decision Logic}       & \multicolumn{1}{c|}{Yes}              & \multicolumn{1}{c|}{Yes}              & \multicolumn{1}{c|}{No}               & \multicolumn{1}{c|}{Yes}              \\ \hline
\multicolumn{1}{|l|}{\textbf{Inputs}}  & \multicolumn{1}{l|}{\textbf{Item}}    & \multicolumn{1}{c|}{\textbf{Stage 1}} & \multicolumn{1}{c|}{\textbf{Stage 2}} & \multicolumn{1}{c|}{\textbf{Stage 3}} & \multicolumn{1}{c|}{\textbf{Stage 4}} \\ \hline
\multicolumn{1}{|l|}{Decision Maker} & \multicolumn{1}{l|}{DM Unit}              & \multicolumn{1}{c|}{Individual}       & \multicolumn{1}{c|}{Group}            & \multicolumn{1}{c|}{Group}       & \multicolumn{1}{c|}{Individual}       \\ \hline
\multicolumn{1}{|l|}{}                        & \multicolumn{1}{l|}{DM Influence}         & \multicolumn{1}{c|}{None}             & \multicolumn{1}{c|}{Seniority/Expert} & \multicolumn{1}{c|}{None}             & \multicolumn{1}{c|}{None}             \\ \hline
\multicolumn{1}{|l|}{Criteria}       & \multicolumn{1}{l|}{Factor Range}         & \multicolumn{1}{c|}{All}              & \multicolumn{1}{c|}{All}              & \multicolumn{1}{c|}{All}              & \multicolumn{1}{c|}{Critical}         \\ \hline
\multicolumn{1}{|l|}{}                        & \multicolumn{1}{l|}{Factor Assessment}    & \multicolumn{1}{c|}{Explicit}         & \multicolumn{1}{c|}{Implicit}         & \multicolumn{1}{c|}{Implicit}         & \multicolumn{1}{c|}{Explicit}         \\ \hline
\multicolumn{1}{|l|}{}                        & \multicolumn{1}{l|}{Rating Type}          & \multicolumn{1}{c|}{Binary}           & \multicolumn{1}{c|}{Likert}           & \multicolumn{1}{c|}{Likert}           & \multicolumn{1}{c|}{Binary}           \\ \hline
\multicolumn{1}{|l|}{}                        & \multicolumn{1}{l|}{Factor Framing}       & \multicolumn{1}{c|}{Ambiguous}        & \multicolumn{1}{c|}{Ambiguous}        & \multicolumn{1}{c|}{Ambiguous}        & \multicolumn{1}{c|}{Ambiguous/Risk}   \\ \hline
\multicolumn{1}{|l|}{}                        & \multicolumn{1}{l|}{Flexible Feasibility} & \multicolumn{1}{c|}{No}               & \multicolumn{1}{c|}{No}               & \multicolumn{1}{c|}{No}               & \multicolumn{1}{c|}{Yes}              \\ \hline
\multicolumn{1}{|l|}{Process}        & \multicolumn{1}{l|}{Reference Point}      & \multicolumn{1}{c|}{Relative}         & \multicolumn{1}{c|}{Stand-alone}      & \multicolumn{1}{c|}{Comparative}      & \multicolumn{1}{c|}{Stand-alone}      \\ \hline
\multicolumn{1}{|l|}{}                        & \multicolumn{1}{l|}{Process Name}         & \multicolumn{1}{c|}{Neutral}          & \multicolumn{1}{c|}{Neutral}          & \multicolumn{1}{c|}{Neutral/Named}    & \multicolumn{1}{c|}{Neutral}          \\ \hline

\end{tabular}
\caption{Stage configuration and outputs}
\label{tab:config_metrics}
}}
\end{table}

The study draws heavily on the expertise of senior DMs with an Australian banking background and familiar with process selection in practice. Across all four stages, 89 potential DMs were approached and 75 agreed to participate. With the exception of one individual, they each had more than ten years' banking experience. All had experience in at least one of the four major Australian banks and 71\% had worked in more than one of them. With the exception of stage 1, all four banks were represented in each stage. DMs had experience in the following functions: operations, technology, risk, projects, operational support, or process excellence (including BPM and process mining experience), and 46\% of them had worked in more than one function. A different group of DMs was selected for each stage, with the exception of the comparative assessment where a subset of DMs from the previous stage was selected. We refer to this group as the Recruitment Pool (RP).

\subsection{Stage 1 -- Base case assessment.}\label{sec:stage_1}

\noindent \textbf{Design \& development.} The primary steps were to: i) derive the CaC; ii) establish the base case impact of the CaC on decision-making; and iii) refine the CaC with the DMs.   

\noindent\textbf{i) Derive the CaC.} 
We leveraged the 23 contextual factors\footnote{The contextual factors are referred to as challenges in the Adams et al., study~\cite{adams2022banks}} identified in~\cite{adams2022banks} as our starting point. We reviewed the factors against the process, organization, and environment contextual dimensions discussed in~\cite{vom2016role} to ensure all dimensions were covered and then mapped them to the selection criteria identified in~\cite{dumas2018fundamentals} to create the CaC.

\noindent \textbf{ii) Base case assessment.} We evaluated the impact of the CaC on process selection relative to our motivating case. We defined a strategic goal, notionally set by the Chief Risk and Compliance Officer (CRCO), to improve Business Process Compliance (BPC) outcomes within twelve months. DMs were informed that the project was fully funded, fully resourced and that the BPM project team had decided to apply process mining techniques to achieve this goal. The objective for the DMs was to determine the suitability of including a given process in the project's scope. 

\noindent To emulate a practical scenario, we recruited two groups of DMs from one bank in the RP and provided them with three variants of a process to assess. The first group was familiar with the payments process -- a \textit{core process} (CP). The second group was familiar with a financial crime process -- a \textit{support process} (SP). We asked each individual DM to complete an online Qualtrics survey to assess the processes.  

\noindent DMs completed the survey over three cycles. For the first cycle, DMs were asked if each process variant was suitable for inclusion, with neither criteria nor context. For the second cycle, they were asked to evaluate each variant against the criteria, but without context, on a ``Yes" / ``No" basis, e.g., ``Is process X strategically important?", before assessing the suitability of each variant. For the third cycle, they were asked to evaluate each variant against each contextual factor and criterion, on a ``Yes" / ``No" basis, before assessing the suitability of the variant. 

\noindent \textbf{iii) Refine the CaC.} A one-hour focus group for each process was conducted after the survey. DMs were taken through the outcomes of the survey and asked to provide feedback. They were also asked to nominate which contextual factors they deemed to be critical, and to identify any missing factors.

\begin{figure}[ht!]
\centering
\includegraphics[width=1.00\textwidth]{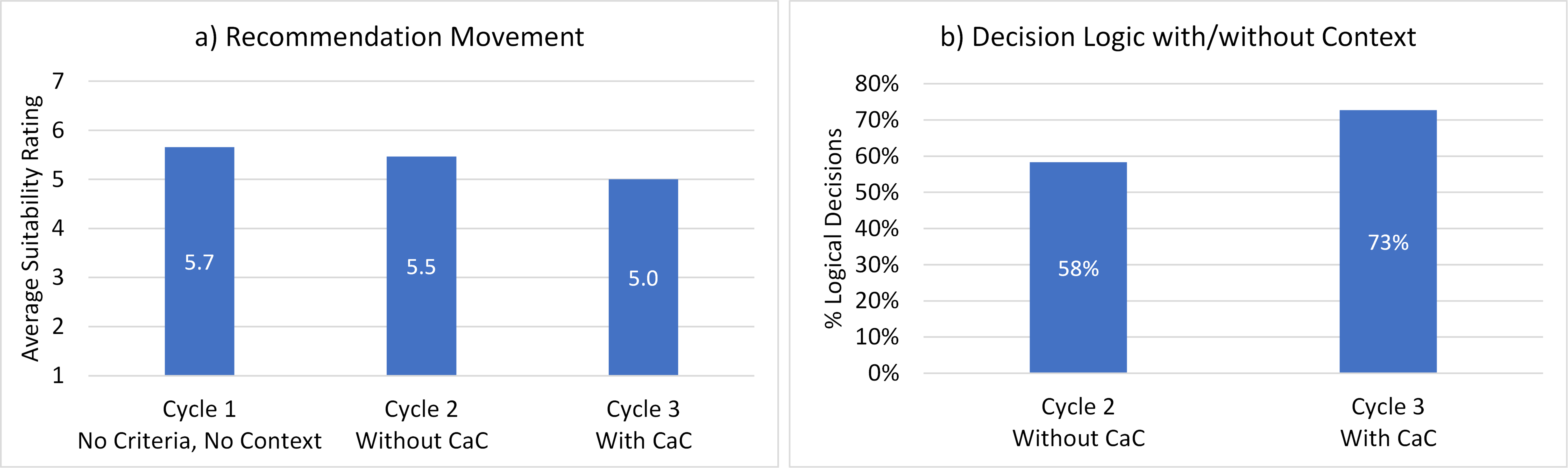}
  \caption{Validate Contextual Factor Outcomes}
  \label{fig:round_1}
\end{figure}

\noindent\textbf{Demonstration -- quantitative.} Survey responses indicated that CaC do influence decision-making (Figure \ref{fig:round_1}a). This addresses RQ1. The pattern was the same for five out of six of the process variants, while one variant was rated the same for each cycle. Introducing CaC also appeared to improve the decision logic (Figure \ref{fig:round_1}b).

\noindent\textbf{Demonstration -- qualitative.}
In practice, process selection is problematic, or as one DM put it: ``It's currently a talk fest until the most senior person in the room makes a call." Another DM alluded to the impact of inconsistent process knowledge, observing that: ``There is a disconnect between the understanding of those making the decision and those with the task of implementing it in practice." DMs referenced ownership, accountability, senior executives changing direction, and aligning competing objectives as challenges. Both groups emphasized access to funding, resources, and expert opinion, particularly with regard to technology, as constraints.

\noindent DMs determined that three of the factors (change-, support- and business model-related factors) were relevant to different criteria, depending on perspective, hence they were split. With process mining selected as the project method, 
the data-related factor was also split into ``data access" and ``common identifier availability". One other feedback-related change was to relabel the \emph{Health} criterion as \emph{Opportunity}. Table \ref{tab:context_criteria} highlights the CaC including the critical factors (italicized), nominated by more than 25\% of DMs.

\begin{table}[]
\centering
{\scriptsize{
\begin{tabular}{|l|l|}
\hline
\textbf{Strategic Importance}        & \textbf{Feasibility (organizational)}    \\ \hline
Process scale                        & \textit{Process ownership}                        \\ \hline
\textit{Extent of regulatory oversight}       & Prior funding                            \\ \hline
Disruptive competitive pressure      & BAU change capacity                      \\ \hline
                                     & Risk culture                             \\ \hline
\textbf{Health/Opportunity}                      & \textit{Stakeholder alignment}                    \\ \hline
\textit{Control quality and effectiveness}    & Business model complexity                 \\ \hline
Process monitoring effectiveness     & Changing business direction              \\ \hline
Level of automation                  & \textbf{Feasibility (data)}              \\ \hline
Number of hand-offs                  & \textit{Access to data}                           \\ \hline
\textit{Level of variation and exceptions}    & Dependency on legacy technology          \\ \hline
\textit{Extent of workarounds and patches}    & Data identifier availability             \\ \hline
Quality of documentation and support & 3rd party involvement                    \\ \hline
Staff knowledge, skills \& experience & Documented requirements                  \\ \hline
\textit{Metrics alignment}                    & Clarity of requirements                  \\ \hline
Change management                    & Requirements conflict and/or duplication \\ \hline
\end{tabular}
\caption{Contextual Factor Mapping to Criteria}
\label{tab:context_criteria}
}}
\end{table}

\noindent \textbf{Observations.} While introducing context appeared to influence the outcome and improve decision logic, it was not possible to validate the quality of the portfolio selection process, given the bespoke, proprietorial nature of each of the process variants assessed. The survey also raised the issue of how DMs familiar with one process, could objectively compare their process to one they were not familiar with. From a CaC perspective, an explicit assessment of all factors made the survey long, and a `Yes" / ``No" response option was too simplistic. For the process input, the variants provided a helpful relative reference point, but, in practice, they will not always be available. Finally, process selection is typically an iterative, group-decision~\cite{elzinga1995business}, whereas this was an individual, once-off assessment and did not reflect ``the most senior person in the room" reality.  

\subsection{Stage 2 -- Individual process assessment.}\label{sec:stage_2}

\noindent \textbf{Design \& development.} To address stage 1 observations, we devised a scenario for DMs to assess one of two hypothetical process descriptions, on a stand-alone basis, as part of a Delphi study. The steps were: i) design the scenario; ii) design and conduct the Delphi study; and iii) assess the ``most senior person in the room" impact.

\noindent \textbf{i) Scenario design.}
We retained the strategic goal from stage 1. We drafted two hypothetical, contextually relevant, process descriptions -- one each for a CP and a SP (Table \ref{tab:process_descr}). The process descriptions comprised a series of qualitative statements with no quantitative facts. The statements were drafted with the contextual factors in mind and indicated the absence or presence of some of these factors. The descriptions were deliberately ambiguous, with both reasons for and against selecting the process for inclusion, and the underlying processes were not readily identifiable. However, each description made it clear, which process was the CP and which process was the SP. To allow us to evaluate the quality of the decision, the descriptions were drafted such that DMs assessing the process with CaC would be unlikely to include the CP process in scope, and likely to include the SP in scope. The process descriptions were validated by the other members of the research team and one industry expert.

\begin{table}[htbp]
  \centering
  {\scriptsize{
  \caption{Hypothetical Process Descriptions}
    \begin{tabular}{|p{24em}|p{24em}|}
    \hline
    
    \multicolumn{1}{|c|}{\textbf{Core Process  (Process A)}} & \multicolumn{1}{c|}{\textbf{Support Process (Process B)}}  \\
    \hline  
    This process is a key driver of current revenue and is anticipated to underpin growth objectives for the foreseeable future. While the process has been subject to a number of regulatory reviews in the past, APRA\textsuperscript{1} now seems comfortable with both current performance and with the roadmap. The process is being transformed as part of a broader program sponsored by the CEO. The program is well-funded, well-resourced and the "minimal viable product" is already in production. What was a clunky process will soon be significantly enhanced with most of the manual steps and workarounds removed. The operations team are currently being re-skilled to handle the small number of exceptions the end-to-end process is expected to produce. The project team has now turned its attention to the client user interface where the team believes they are behind the competition and there is significant scope to improve the client experience.\newline{}\newline{}Risk \& compliance are working with the various business unit heads and 3rd party ecosystem providers to interpret and refine related obligations and policies as well as resolve conflicting objectives, which is taking longer than expected. The quality of the data the process will produce will enable ongoing performance improvement and make performance truly transparent. However, delivering the data may be pushed into the next financial year, given the focus on getting customer-facing functionality out to market and broader bank constraints around cost. \newline{} \newline{} \textsuperscript{1} APRA is an Australian banking regulator.
    & While this process does not generate revenue directly, many revenue-producing processes depend on it. As such, it is not directly visible to customers unless it fails, which it is doing more regularly. In the past it has been considered a “hygiene” process, although in more recent times it has come under greater regulatory scrutiny as there appear to be some control gaps. It is also an area capturing the public's attention as a slew of FinTechs try to disrupt this area. Operators must learn to accommodate a large number of exceptions. The quality of training and support material is poor which means operator time to competency is measured in months not weeks. It is a relatively manual process with lots of workarounds and, given this complexity, operators continue to make routine errors. Re-platforming is still a long way off. \newline{}\newline{}A recent restructure consolidated a number of disparate teams into one business unit and a new leader, well-respected across the organization, was appointed. With the re-structure complete, the combined leadership team is now very experienced. It has a clear mandate, clear obligations and clear requirements. Their revised strategy has been well-received by senior stakeholders, although a number of front-line managers have been vocal that the organization should be prioritizing revenue generating- and customer experience-oriented projects. From Marketing's perspective there have been some issues trying to migrate the data to the CRM platform, however, at an operational level the business has access to detailed transactional data and their current event reporting can identify issues promptly and precisely in whichever system the error occurred. \\
    \hline
    \end{tabular}%
  \label{tab:process_descr}%
  }}
\end{table}%

\noindent\textbf{ii) Delphi study.}
For this stage we selected a Delphi study method, which is appropriate where there is limited information and expert insight is required to reach a group-based, consensus decision~\cite{beiderbeck2021preparing}. Moreover, it is conducted on an anonymous basis, ensuring every voice is equally heard, including those of the experts and the more senior team members. It also provided flexibility to vary the rating approach (an implicit not explicit factor assessment and Likert scales not binary responses), as well as assessing a stand-alone process.

\noindent Four anonymous, cross-functional groups of six DMs were randomly selected from the RP. Each group was assigned to a process-criteria question set. Two sets of criteria questions were drafted. One set asked DMs to assess their process relative to each criterion, without context, on a seven-Point Likert scale. The second set was the same except that each criterion question incorporated the full list of the mapped contextual factors, i.e., an implicit assessment of the CaC.  After rating the criteria, DMs assessed their ``likelihood to recommend" the process on a six-point Likert scale and were asked to justify each assessment. Responses were captured in a Qualtrics online survey. Each group had at least one DM familiar with process mining. The study ran for three weekly cycles and, at the end of each cycle, DMs were provided with the verbatim comments and a frequency chart indicating how group members had assessed their process.

\noindent \textbf{iii) Seniority response -- the CRCO email.} We concluded the stage by sending DMs an email purporting to be the CRCO's views, to assess the impact of the ``most senior person in the room" (cycle 4). For the CP, the email emphasized the facts in the process description that supported not recommending the process. For the SP, the email challenged the facts in the description, adopting a ``wait and see" tone. DMs were asked to respond and re-assess their likelihood to recommend the relevant process for inclusion.

\begin{figure}[ht!]
\centering
\includegraphics[width=1.00\textwidth]{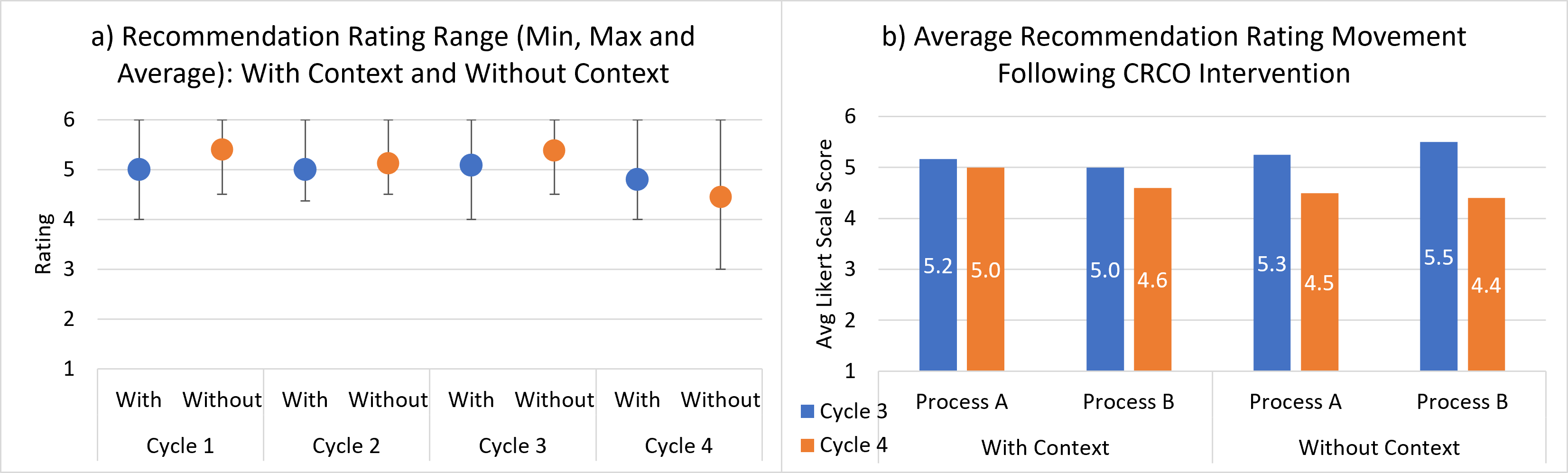}
  \caption{Individual Assessment outcomes}
  \label{fig:round_2}
\end{figure}

\noindent\textbf{Demonstration -- quantitative.}
The outcomes, for the most part, were confounding. During the first three cycles, there was little movement in likelihood of recommendation for any of the groups (Figure \ref{fig:round_2}a). While some DMs did change their mind during the three cycles, the range was limited, with only 20\% of decisions moving by more than one Likert step. All DMs would at least ``Possibly" recommend their process for inclusion, and 83\% of DMs would either ``Probably" or ``Definitely" recommend their process for inclusion.  In summary, the SP was preferred from an \emph{Opportunity} perspective, but for the other two criteria and the recommendation, the CP was slightly ahead. 

\noindent In terms of differences between groups based on the question set they were provided with, groups without the CaC tended to rate slightly higher and with lower variability on all decisions with the exception of cycle 4 for the recommendation rating. The ratings for both the criteria and the recommendation were high, hence the decision logic was also high (86\%). However, for the  groups with CaC it was lower than those without CaC (83\% versus 89\%), contradicting stage 1 findings.

\noindent The expert voice was not heard. The CP description made clear that data was not going to be available for another year -- an obvious feasibility impediment for a process mining project. This was noted in the rating of the process mining expert in the relevant CaC group, but the rating had no impact on other members of the group over the three cycles with a gap of 2.8 Likert points.

\noindent The most marked change in this stage followed the CRCO's email (Figure \ref{fig:round_2}b). All groups rated their likelihood to recommend lower, the groups without CaC by the most (0.9 versus 0.3) and the SP more than the CP (0.8 versus 0.5). 

\noindent\textbf{Demonstration -- qualitative.}
DMs expressed difficulty assessing the processes in isolation with neither a reference process to compare against nor a process name. For the CP, its revenue-generating attribute was interpreted as a key driver of \emph{Importance}, despite the compliance-oriented nature of the project goal. The CEO's sponsorship of a transformation project was viewed favorably. DMs discounted the regulator's comfort level with the CP, assuming that:  ``Once on the radar, always on the radar". They also assumed that both the conflicting objectives and data-related issues could be addressed. Referring to the SP as a ``hygiene" process, diminished its value in the eyes of DMs, although they recognized the potential \emph{Opportunity}. There was a mixed reaction to the new management team - some saw it as a positive, others wanted to wait and see. From a \emph{Feasibility} perspective, DMs were concerned about the lack of investment, the difficulties of addressing the legacy issues, and suggested funding would be difficult (even though they had been informed that the project was fully funded).

\noindent In response to the CRCO email, the CP DMs expressed disappointment but acceptance. They saw it as being driven by the regulator's position and suggested the regulator's confidence may be misplaced. They encouraged the CRCO to push for the data, at least for a pilot project. For the SP, the tone was more negative and despondent. DMs were dismissive of the CRCO, expressing concerns that they were burying the issues: ``This is a newspaper headline waiting to happen", running a hidden agenda, and that the way forward was to get hard facts.

\noindent \textbf{Observations.} While the decision logic was high, the decision quality did not match the intended design -- DMs with CaC rated the CP slightly higher than DMs with CaC rated the SP. Our primary design consideration for the next stage was to gather more feedback from DMs to help interpret these results. Additional points to consider were the feedback regarding the lack of reference point.

\subsection{Stage 3 -- Comparative process assessment.}\label{sec:stage_3} 
\noindent \textbf{Design \& development.} Considering the outcome from stage 2, there were two steps for this stage: i) capture additional feedback on stage 2; and ii) understand the impact of a comparative assessment and process naming as a reference point. 

\noindent \textbf{i) Stage 2 follow-up.}
Given the exploratory nature of the first step, we chose to conduct a one-hour, semi-structured focus group to capture data and invited volunteers from Stage 2 to participate. Seven DMs agreed. We opened the focus group with a discussion of the outcomes from stage 2. 

\noindent \textbf{ii) Process comparison.}
Continuing the stage 2 scenario, we ran two comparative assessment cycles during the focus group. For the first, the processes were referred to by neutral labels -- Process A (the CP) and Process B (The SP). DMs were asked to provide a comparative rating for each criterion on a seven-point Likert scale, e.g., ``Process A is significantly more feasible", and which process they were more likely to recommend for inclusion in the scope, e.g., ``Process B: slightly more likely". This also allowed us to evaluate the issue of process familiarity as DMs were more familiar with the process they had assessed in stage 3 than the alternate process. For the second cycle, the  questions were the same, but DMs were informed that Process A was loosely based on a mortgage process and Process B was loosely based on a KYC process.  

\begin{figure}[ht!]
\centering
\includegraphics[width=1.00\textwidth]{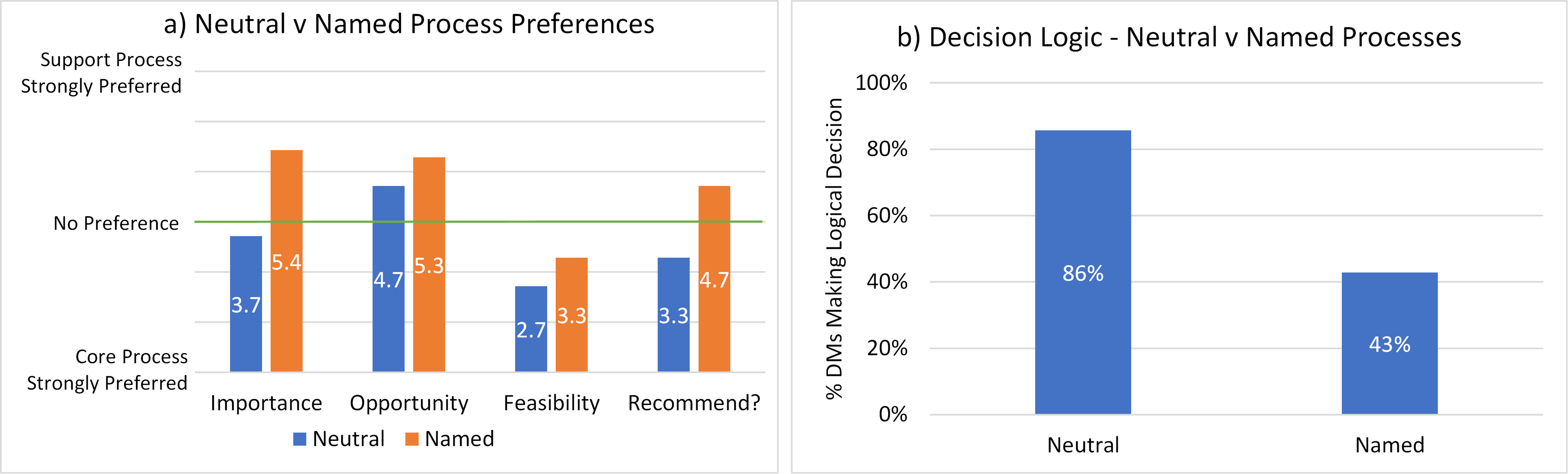}
  \caption{Comparative Assessment outcomes}
  \label{fig:round_3}
\end{figure}

\noindent\textbf{Demonstration -- quantitative.} The chart in Figure \ref{fig:round_3}a shows the extent of preference switching by enabling a comparative assessment and naming the processes, with Process B (KYC) clearly favored in three of the four categories as opposed to one on a name-neutral basis. The justification for the switch was that if you failed to comply with KYC compliance requirements: ``You go to jail." However, DMs commented that a decade ago, before the current wave of financial crime penalties had been imposed, a similar response would have been extremely unlikely. The decision logic deteriorated when the processes were named (Figure \ref{fig:round_3}b). 

\noindent  While DMs suggested they struggled to rate the processes in stage 2 without a reference point, there was little evidence of this. Comparing ratings between stage 2 and stage 3, cycle 1, there was no difference in the \emph{Importance} and \emph{Opportunity} ratings and only minor rating differences in \emph{Feasibility} and the \emph{Likelihood to Recommend} assessments.

\noindent\textbf{Demonstration -- qualitative.} Contextually DM comments raised a number of points with regards to the Delphi study. DMs stated that they anchored their position in cycle 1 and, in the absence of new facts, resolved not to change. Views of the other experts in the group were not considered new facts. They also stressed that the CP appeared to be positioned as an opportunity -- which aligns to the criterion label -- and the SP was positioned as a threat. The group emphasized the likelihood of securing funding is greatly increased if the process is seen as a revenue generating opportunity, reinforcing the point made in~\cite{malinova2018identifying}. The group commented on the subjective nature of the process in practice: ``There's subjectivity, even when you do have the data". They also highlighted the political reality of the process in practice, where securing funding for your project is seen as a pathway to an improved individual performance outcome. As such, it is common practice to ``dress up" an investment case or change the narrative if the first review wasn't received positively. It is also a process framed by the financial calendar: "We work in one-year blocks, that's how we are rewarded." 

\noindent \textbf{Observations.} While in cycle 2 the decision quality aligned to our design intent (the SP was preferred to the CP), the decision logic was below 50\%. This stage did, however, provide some answers to the underwhelming outcome from stage 2. It is not just the CaC that provide the context for the process, the other process inputs also influence the context. The impact of DM personal bias (particularly with regards to conflicting agenda), the importance of framing the goal, and the impact of naming the process, all played a significant role in the portfolio selection process. One other point of interest was the importance DMs placed on the need for \emph{Feasibility} flexibility to overcome potential roadblocks within the operating rhythm time frame of the organization.

\subsection{Stage 4 -- Pre-screening assessment.}\label{sec:stage_4} 
\noindent \textbf{Design \& development.} Continuing with the stage 2 scenario, we consolidated the insights from previous stages into the design of the final stage -- a survey-only approach. There were two steps: i) survey design; and ii) cycle design. 

\noindent \textbf{i) Design survey.}
The research method for this stage was an online,  Qualtrics survey. DMs were randomly assigned to one of the two processes. As a pre-screening stage, the underlying intent was to ``screen-out" unsuitable processes. Hence, for the survey design, we focused on the critical contextual factors identified in stage 1 and reverted to an explicit assessment on a ``Yes" / ``No" basis. On review, we consolidated four of the factors associated with the \emph{Opportunity} criterion to reduce the risk of isolation effects~\cite{kah1979prospect}. Hence, there were five questions in the survey, one each for the \emph{Value} criteria (\emph{Importance} and \emph{Opportunity}) and three for the \emph{Feasibility} criteria. DMs were also asked what information was missing and what questions they would have asked. 

\noindent If the DM responded ``No" to a \emph{Feasibility} criterion question, a follow-up question was displayed, asking if they thought the issue could be addressed in a reasonable time frame. If the issue could be addressed, the factor was considered a ``potential roadblock", if not, it was considered a ``roadblock". A positive response to either of the \emph{Value} criteria was considered a ``green flag".

\noindent \textbf{ii) Cycle design.}
There were two cycles completed by 20 DMs each. For the first cycle, the \emph{Value} criteria were labeled \emph{Importance} and \emph{Opportunity}. For the second cycle, the criteria were re-labeled as \emph{Risk} criteria to \emph{Impact} and \emph{Likelihood} respectively i.e., in risk language, and the questions were also re-framed.  

\begin{figure}[ht!]
\centering
\includegraphics[width=1.00\textwidth]{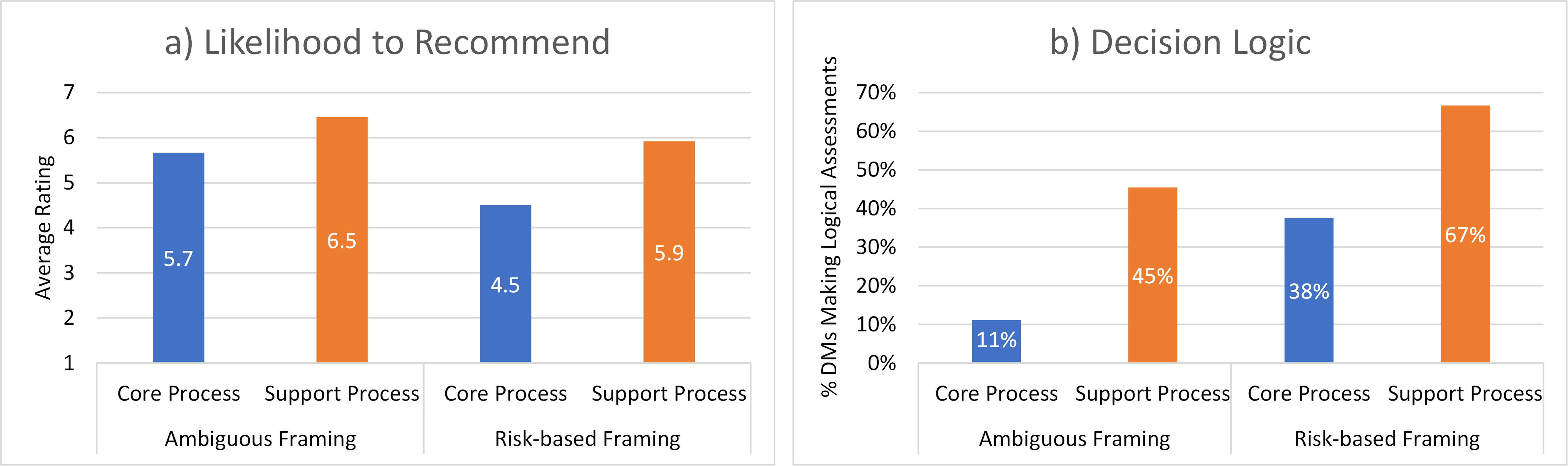}
  \caption{Pre-screening outcomes}
  \label{fig:round_4}
\end{figure}

\noindent \textbf{Demonstration -- quantitative.} In cycle 1, the SP was favored (Figure \ref{fig:round_4}a). The driver was \emph{Feasibility} where, on average each DM rating the CP assigned 0.9 roadblocks, for the SP the corresponding number was 0.2.

\noindent  Only 11\% of CP ratings and 45\% of SP ratings were considered logical in cycle 1. However, framing the \emph{Value} criteria as \emph{Risk} criteria widened the gap and improved the logic. For the CP the number of ``green flags" fell from 1.7 to 0.8 per assessment, and although the \emph{Feasibility} criterion did not change, the number of  ``roadblocks" increased from 0.9 to 1.4. For the SP the number of ``potential roadblocks" fell from 0.9 to 0.5, otherwise there was no change. The likelihood to recommend also changed from 5.7 to 4.5 for the CP and to a lesser extent for the SP (6.5 to 5.9). The decision logic improved following re-framing for both processes, albeit for the CP it was still below 40\%. 

\noindent At a more detailed level, the number of ``potential roadblocks" that DMs believed could be addressed without impacting the project was significantly higher for the CP than the SP -- four times more likely for \emph{Aligned Objectives}, twelve times more likely for \emph{Data Availability} and five times more likely for \emph{Accountability}. 

\noindent \textbf{Demonstration -- qualitative.}
In terms of qualitative comments, DMs wanted more data -- volumes, revenue, number of incidents, control failures, risk assessments, and common process metrics, e.g., timeliness, quality and cost. DMs expressed frustration that there was a limited set of factors to evaluate. They also wanted to see the other alternatives that were being considered, as well as understand the broader operating environment. Finally, one DM commented that unless there is significant regulatory pressure, a compliance-related project will not be funded without considering other value drivers, e.g., customer experience, revenue growth or productivity and cost reduction.

\noindent \textbf{Observations.} Framing the context consistently, with a limited number of critical factors, allowing for roadblocks to be addressed and evaluated explicitly on a ``Yes" / ``No" basis, generated a better quality, more logical outcome. 
\section{Discussion}\label{sec:discussion}
In this section we discuss the implications of our study and provide a set of guidelines for both BPM practitioners and academics. 

On the one hand, by stage 4, we satisfied our proposition that a DM with CaC would make better, more logical decisions -- DMs showed a strong preference for the support process over the core process. However, the principal question emanating from stages 2-4 was: ``Given a compliance-oriented goal, why are DMs prepared to support a process that is of limited concern to the regulator, has no acknowledged health concerns, is inhibited by a lack of data and where stakeholders haven't agreed on their objectives?" 

This process is both politically charged and prone to decision-making bias -- or from Mel{\~a}o and Pidd's perspective, it is ``process as a social construct"~\cite{melao2000conceptual}. Where the organizational reward system encourages people to compete for funding, it is not surprising to hear DMs discuss ``dressing up the case" or ``changing the narrative" to get their bid approved, and that the investment process is heavily oversubscribed. While context matters, how it is configured is critically important. Hence, designing a context-aware selection process is not as simple as evaluating process alternatives against a list of contextual factors. The project goal, DMs and the process alternatives also influence the context, and not all configurations lead to better quality, more logical decisions. How, where and when the factors are configured will determine whether the desired outcome is achieved. Below, we discuss five configuration guidelines to assist.

\begin{figure}[ht!]
\centering
\includegraphics[width=1.00\textwidth]{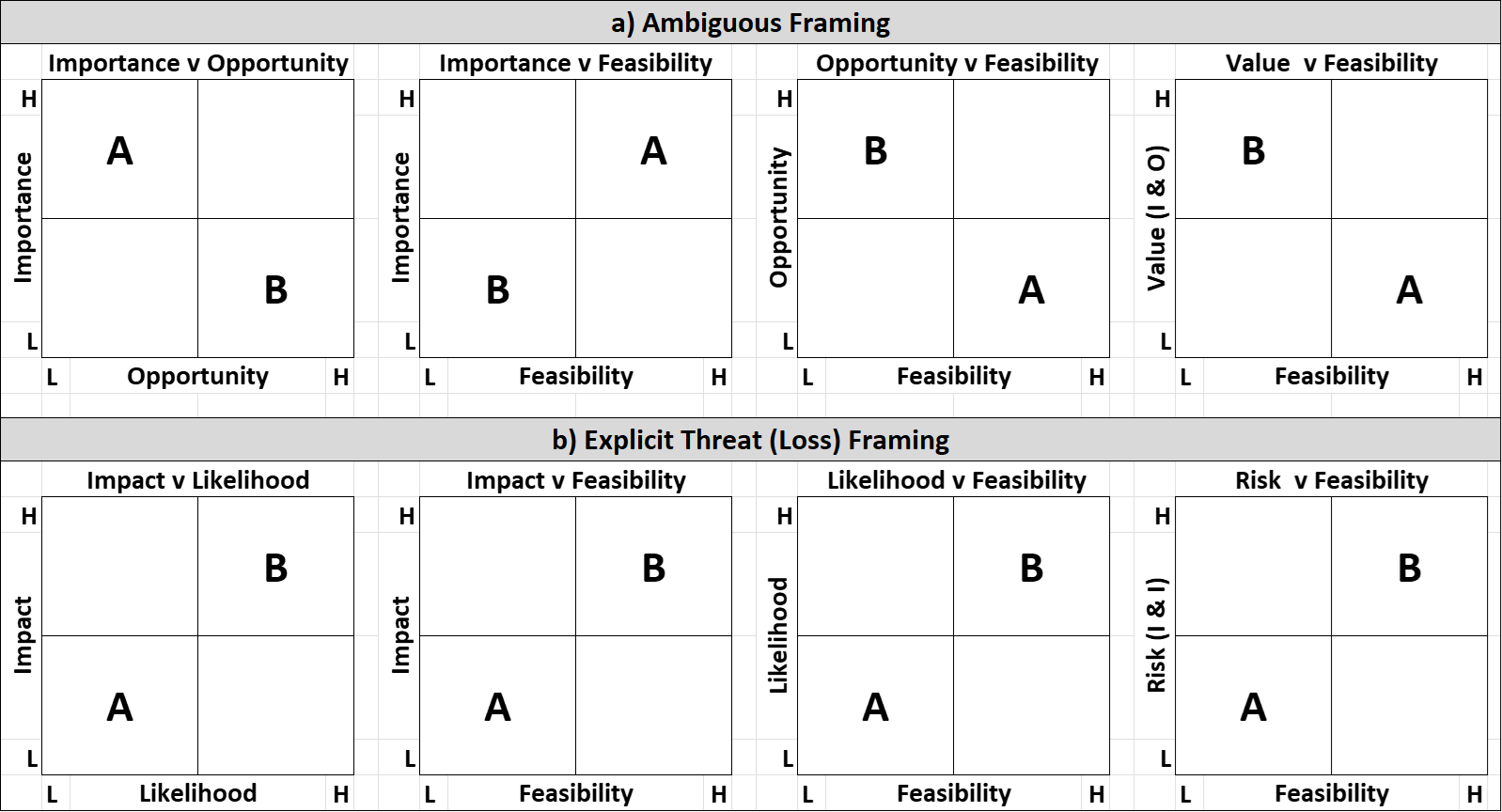}
  \caption{Framing Impact: Opportunity v Threat}
  \label{fig:framing}
\end{figure}

\textbf{Guideline 1 -- Frame the context.}
In stage 3, DMs commented that Process A presented as a ``low-hanging fruit" opportunity, whereas Process B presented as a potential threat. The CP was preferred not only because DMs saw it as easier, but, given this positioning, they considered the scope of the project to cover other opportunities such as customer experience and productivity benefits. Reframing the criteria labels from \emph{Importance} and \emph{Opportunity} to \emph{Impact} and \emph{Likelihood} changed the outcome. The rating difference in stage 4 widened from 0.8 in favor of the SP to  1.6 (Figure \ref{fig:round_4}a), the decision logic of the CP quadrupled and that of the SP doubled (Figure \ref{fig:round_4}b). In stages 2 to 4a, the framing was ambiguous. The project goal was compliance oriented (implying threat and loss) but seeking to improve outcomes (implying opportunity and wins). This ambiguity led to an inconclusive selection outcome, highlighted in Figure \ref{fig:framing}a. When framed explicitly as a risk-based decision (threat), the pattern is consistent and the choice is far more obvious (Figure \ref{fig:framing}b).

Framing the decision as an opportunity or threat and by extension the fourfold pattern considered in Prospect Theory~\cite{kah1979prospect}, sets the tone for describing the project goal, selecting the criteria and identifying the contextual factors cascading through the language and labeling applied across the process. 

\textbf{Guideline 2 -- Level the playing field.} In a subjective process, context is in the eye of the beholder, where inconsistent familiarity with the alternatives, conflicting agendas, and personal bias create tension in the process. There are several design options to mitigate this risk at the outset: i) early access to expert opinion; ii) a shared, common fact base; iii) neutral language and labeling; and iv) neutral process names.  

Each group in stage 2 had at least one DM familiar with process mining, and while they commented that data was an issue for the CP and their \emph{Feasibility} rating reflected this concern, their voice was not heard. The reasoning was, according to one DM in stage 3, that they were already anchored to their decision. Where new concepts, methods, and/or technologies are introduced, equal access to expert opinion is required for all DMs prior to pre-screening.

The response to the SP CRCO email, highlights the importance of a shared, common fact base. DMs assumed the CRCO was working to a different agenda and their respective assessment fell by 13\%, whereas the CP assessment fell by 8\% (Figure \ref{fig:round_2}b). The CRCO email also highlighted the importance of establishing the sponsor's position at the start of the process.

The third option is the choice of language for labeling. In all stages, DMs suggested the performance metrics they would like to inform their decision. This included metric names associated with a specific process, e.g., STP (straight-through-processing) is a common payments metric. Using this metric would reveal the process identity and potentially introduce bias. Relabeling the metric relative to its intent, e.g., ``Automation Level", which can be applied to other processes, would appear to be more appropriate.

As stage 3 highlighted, the name of the process is particularly influential. Deidentifying the processes, and describing them in context-aware, neutral terms, mitigates the effects of familiarity bias. In stage 3 the decision logic fell from 86\% to 43\% when the processes were named. 

Aspects of this may not appear to be practically acceptable, particularly not naming the process. However, as part of a shadow control process, it can act as a check on the actual selection process. Drafting neutrally named, context-aware process descriptions, with neutral labels and fact bases, prepared by an independent team, e.g., a centralized BPM team or consultants, creates a level playing field for DMs. Asking DMs to then pre-screen, individually assess and comparatively assess processes prior to running the traditional process will serve as a control point.

\textbf{Guideline 3 -- Offer a lifeline.} Stage 2 highlighted the limitations of assessing feasibility factors implicitly. DMs recognized feasibility constraints but suggested that they could be overcome. What wasn't clear, was how long this would take and whether it was realistic. This was addressed in stage 4, where 43\% of the \emph{Feasibility} assessments were initially rated as not feasible, but of these 56\% were deemed to be addressable without impacting the duration of the project. The key point here is to offer DMs a lifeline and not to exclude a process before determining whether roadblocks or a weak value case can be addressed. 

\textbf{Guideline 4 -- Choose horses for courses.} The integrated framework in~\cite{archer1999integrated} is essentially an iterative filtering process. Not all contextual factors are relevant at each stage and how they are presented should vary: ``While it would be easier to have everything on a binary checklist, the reality is it doesn't always work". A binary ``Yes" / ``No" approach in pre-screening, assuming the reasoning is clear to DMs, is an effective way to rapidly filter an oversubscribed portfolio (stage 4, cycle 2). At the individual analysis stage, a more detailed and nuanced profiling of a broader set of factors is required. This must provide suitable organizational reference points, such as parameters for ``Low", ``Medium" and ``High" ratings, to enable DMs to calibrate, visualize and self-screen their process prior to the comparative assessment. Finally, at the comparative stage, rationalizing the factors to those that discriminate in line with the isolation effect~\cite{kah1979prospect} and scoring, whether it is a simple RAG rating or applying the more sophisticated MCDA tools, provides a further-level of fine tuning. 

Striking a balance between providing too much context versus too little context aligns with Proposition 4 in~\cite{archer1999integrated}, however there appears to be little research on how to do this. This study presented one approach to selecting the critical contextual factors for pre-screening (stage 1). Stage 2 expanded to the full list of factors and stage 4 collapsed the four critical factors mapped to the \emph{Opportunity/Likelihood} criterion. 

\textbf{Guideline 5 -- Control the room.} 
Process selection is a group decision-making process~\cite{elzinga1995business}, but the groups are small, and, even then, not all voices are heard. The presence or absence of a single, influential DM, for a specific stage can impact the outcome. Perhaps unsurprisingly, the DMs with a risk and compliance background in this exercise were more likely to assess both the criteria and recommendation higher than the other functions. Hence, designing a selection process to ensure consistent, balanced representation throughout the process and configuring contextual factors to mitigate these effects is critical.   

\subsection{Limitations}
There are a number of limitations associated with this study. There are many possible contextual configurations. We assessed a sample of those we believed would carry the most weight. Our iterative approach meant that comparing data points between stages was not always possible. The number of participants assessing a specific configuration at each stage was relatively small -- between four and ten participants. In the Delphi study one group was reduced to four DMs after the first cycle due to participant drop out. To try and avoid familiarity bias, each stage, with the exception of stage 3, drew on a new group of participants. In practice, a group would work through each stage of the selection process and would adapt their decision-making to the stage. Finally, this study was undertaken in an Australian banking context, while we have no reason to believe the findings cannot be generalized, further research is required to test its validity in other jurisdictions.
\section{Conclusion}\label{sec:conclusion}
Context-aware BPM is a developing area of research. In this study our objective was to understand the impact of context on process selection -- a BPM lifecycle stage with known pitfalls. Our intent was to determine whether a DM applying CaC would make a better quality, more logical decision. Our study design iterated through four stages, requiring a functional mix of DMs, drawn from the banking industry, to indicate their likelihood to recommend a process for selection, based on qualitative, subjective criteria only. The study validated our proposition and also highlighted that process selection is a subjective process, prone to bias and is politically charged. In practice, the goal for DMs is to get their process approved, even if that means ``dressing up" the case or changing the process narrative to fit the context. 

Context matters, and all inputs influence the context, but how a context-aware selection process is configured will determine its effectiveness. In this study we identified five guidelines that will improve the quality and logic of the outcome. Clear framing, with a consistent, balanced group of Decision Makers, on a level playing field, with an option to address potential roadblocks, and context-aware criteria configured to the relevant process stage, is a step in the right direction. 

While there are limitations to the study, the results are promising in terms of both the decision-quality and logic that we achieved. More research is required to determine how to select the right mix of contextual factors at each stage of the process. The next stage of our work is to apply these guidelines holistically in a field setting.

\smallskip\noindent\textbf{Acknowledgments:} We thank the participants who generously gave their time and shared their thoughts to help produce this paper.

\bibliographystyle{abbrv}
\bibliography{QUT,Banking}

\begin{thebibliography}{10}

\bibitem{adams2022banks}
N.~Adams, A.~Augusto, M.~Davern, and M.~L. Rosa.
\newblock Why do banks find business process compliance so challenging? an
  australian perspective.
\newblock In {\em BPM}, pages 3--20. Springer, 2022.

\bibitem{archer1999integrated}
N.~P. Archer and F.~Ghasemzadeh.
\newblock An integrated framework for project portfolio selection.
\newblock {\em International Journal of Project Management}, 17(4):207--216,
  1999.

\bibitem{becker1999identifying}
J.~Becker, C.~v. Uthmann, M.~zur Muhlen, and M.~Rosemann.
\newblock Identifying the workflow potential of business processes.
\newblock In {\em HICSS}, pages 10--pp. IEEE, 1999.

\bibitem{beiderbeck2021preparing}
D.~Beiderbeck, N.~Frevel, A.~Heiko, S.~L. Schmidt, and V.~M. Schweitzer.
\newblock Preparing, conducting, and analyzing delphi surveys:
  Cross-disciplinary practices, new directions, and advancements.
\newblock {\em MethodsX}, 8:101401, 2021.

\bibitem{benner2003exploitation}
M.~J. Benner and M.~L. Tushman.
\newblock Exploitation, exploration, and process management: The productivity
  dilemma revisited.
\newblock {\em Academy of management review}, 28(2):238--256, 2003.

\bibitem{cho2011study}
C.~Cho and S.~Lee.
\newblock A study on process evaluation and selection model for business
  process management.
\newblock {\em Expert Systems with Applications}, 38(5):6339--6350, 2011.

\bibitem{darmani2013business}
A.~Darmani and P.~Hanafizadeh.
\newblock Business process portfolio selection in re-engineering projects.
\newblock {\em Business Process Management Journal}, 2013.

\bibitem{davenport1993process}
T.~H. Davenport.
\newblock {\em Process innovation: reengineering work through information
  technology}.
\newblock Harvard Business Press, 1993.

\bibitem{denner2018context}
M.-S. Denner, M.~R{\"o}glinger, T.~Schmiedel, K.~Stelzl, and C.~Wehking.
\newblock How context-aware are extant bpm methods?-development of an
  assessment scheme.
\newblock In {\em BPM}, pages 480--495. Springer, 2018.

\bibitem{druckman2001evaluating}
J.~N. Druckman.
\newblock Evaluating framing effects.
\newblock {\em Journal of economic psychology}, 22(1):91--101, 2001.

\bibitem{dumas2018fundamentals}
M.~Dumas, M.~La~Rosa, J.~Mendling, and H.~A. Reijers.
\newblock {\em Fundamentals of business process management (Second Edition)}.
\newblock Springer, 2018.

\bibitem{edwards1996prospect}
K.~D. Edwards.
\newblock Prospect theory: A literature review.
\newblock {\em International review of financial analysis}, 5(1):19--38, 1996.

\bibitem{elzinga1995business}
D.~J. Elzinga, T.~Horak, C.-Y. Lee, and C.~Bruner.
\newblock Business process management: survey and methodology.
\newblock {\em IEEE transactions on engineering management}, 42(2):119--128,
  1995.

\bibitem{hakim2016fuzzy}
A.~Hakim, M.~Gheitasi, and F.~Soltani.
\newblock Fuzzy model on selecting processes in business process reengineering.
\newblock {\em Business Process Management Journal}, 2016.

\bibitem{kah1979prospect}
D.~Kahneman and A.~Tversky.
\newblock Prospect theory: An analysis of decision under risk.
\newblock {\em Econometrica}, 47(2):363--391, 1979.

\bibitem{kettinger1997business}
W.~J. Kettinger, J.~T. Teng, and S.~Guha.
\newblock Business process change: a study of methodologies, techniques, and
  tools.
\newblock {\em MIS quarterly}, pages 55--80, 1997.

\bibitem{leshob2018towards}
A.~Leshob, A.~Bourgouin, and L.~Renard.
\newblock Towards a process analysis approach to adopt robotic process
  automation.
\newblock In {\em ICEBE}, pages 46--53. IEEE, 2018.

\bibitem{levy1996loss}
J.~S. Levy.
\newblock Loss aversion, framing, and bargaining: The implications of prospect
  theory for international conflict.
\newblock {\em International Political Science Review}, 17(2):179--195, 1996.

\bibitem{list2004neoclassical}
J.~A. List.
\newblock Neoclassical theory versus prospect theory: Evidence from the
  marketplace.
\newblock {\em Econometrica}, 72(2):615--625, 2004.

\bibitem{malinova2018identifying}
M.~Malinova and J.~Mendling.
\newblock Identifying do’s and don’ts using the integrated business process
  management framework.
\newblock {\em Business Process Management Journal}, 2018.

\bibitem{melao2000conceptual}
N.~Mel{\~a}o and M.~Pidd.
\newblock A conceptual framework for understanding business processes and
  business process modelling.
\newblock {\em Information systems journal}, 10(2):105--129, 2000.

\bibitem{rosemann2006context}
M.~Rosemann and J.~Recker.
\newblock Context-aware process design: Exploring the extrinsic drivers for
  process flexibility.
\newblock In {\em Proceedings of the Workshops and Doctoral Consortium}, pages
  149--158. Namur University Press, 2006.

\bibitem{rosemann2008contextualisation}
M.~Rosemann, J.~Recker, and C.~Flender.
\newblock Contextualisation of business processes.
\newblock {\em International Journal of Business Process Integration and
  Management}, 3(1):47--60, 2008.

\bibitem{saaty2008decision}
T.~L. Saaty.
\newblock Decision making with the analytic hierarchy process.
\newblock {\em International journal of services sciences}, 1(1):83--98, 2008.

\bibitem{terres2010selection}
L.~D. Terres, J.~A.~R. Nt, and J.~M. de~Souza.
\newblock Selection of business process for autonomic automation.
\newblock In {\em EDOC}, pages 237--246. IEEE, 2010.

\bibitem{trkman2010critical}
P.~Trkman.
\newblock The critical success factors of business process management.
\newblock {\em International journal of information management},
  30(2):125--134, 2010.

\bibitem{tversky1992advances}
A.~Tversky and D.~Kahneman.
\newblock Advances in prospect theory: Cumulative representation of
  uncertainty.
\newblock {\em Journal of Risk and uncertainty}, 5:297--323, 1992.

\bibitem{vom2021context}
J.~vom Brocke, M.-S. Baier, T.~Schmiedel, K.~Stelzl, M.~R{\"o}glinger, and
  C.~Wehking.
\newblock Context-aware business process management.
\newblock {\em Business \& Information Systems Engineering}, 63(5):533--550,
  2021.

\bibitem{vom2018frameworks}
J.~vom Brocke and J.~Mendling.
\newblock Frameworks for business process management: a taxonomy for business
  process management cases.
\newblock {\em Business process management cases: digital innovation and
  business transformation in practice}, pages 1--17, 2018.

\bibitem{vom2014ten}
J.~Vom~Brocke, T.~Schmiedel, J.~Recker, P.~Trkman, W.~Mertens, and S.~Viaene.
\newblock Ten principles of good business process management.
\newblock {\em Business process management journal}, 2014.

\bibitem{vom2016role}
J.~vom Brocke, S.~Zelt, and T.~Schmiedel.
\newblock On the role of context in business process management.
\newblock {\em International Journal of Information Management},
  36(3):486--495, 2016.

\bibitem{wanner2019process}
J.~Wanner, A.~Hofmann, M.~Fischer, F.~Imgrund, C.~Janiesch, and
  J.~Geyer-Klingeberg.
\newblock Process selection in rpa projects-towards a quantifiable method of
  decision making.
\newblock 2019.

\bibitem{wkatrobski2019generalised}
J.~W{a}tr{\'o}bski, J.~Jankowski, P.~Ziemba, A.~Karczmarczyk, and M.~Zio{\l}o.
\newblock Generalised framework for multi-criteria method selection.
\newblock {\em Omega}, 86:107--124, 2019.

\end{thebibliography}

\end{document}